\documentclass[a4paper,12pt]{iopart}
\usepackage[utf8]{inputenc}
\usepackage{graphicx}
\usepackage{iopams}
\usepackage{bm}
\usepackage{graphicx} 
\def\<{\langle}
\def\>{\rangle}

\begin{document}

\title[Dynamics and instantaneous normal modes]{Dynamics and instantaneous normal modes in a liquid with density anomalies}
\author{M P Ciamarra$^{1,2}$ and P Sollich$^3$}

\address{
$^1$ Division of Physics and Applied Physics, School of
Physical and Mathematical Sciences, Nanyang Technological University, Singapore}
\address{
$^2$  CNR--SPIN, Dipartimento di Scienze Fisiche,
Universit\`a di Napoli Federico II, I-80126, Napoli, Italy
}
\address{
$^3$  King’s College London, Department of Mathematics, Strand, London WC2R 2LS, United Kingdom
}

\begin{abstract}
We investigate the relation between the dynamical features of a supercooled liquid 
and those of its potential energy landscape,
focusing on a model liquid with density anomalies.
We consider, at fixed temperature, pairs of state points with different density but the same diffusion constant, 
and find that surprisingly they have identical dynamical features
at all length and time scales. This is shown by the collapse of their mean square displacements
and of their self--intermediate scattering functions at different wavevectors.
We then investigate how the features of the energy landscape change with density,
and establish that state points with equal diffusion constant have different landscapes. 
In particular, we find a correlation between the fraction of instantaneous normal modes connecting 
different energy minima and the diffusion constant, but unlike in other
systems these two quantities are not in one--to--one correspondence with each other, showing that additional landscape features must be relevant in determining the 
diffusion constant.
\end{abstract}

\pacs{61.43.Fs,66.10.Cb,61.20.Lc}

\submitto{jpcm}
\maketitle
\section{Introduction}
Particle motion in glassy systems is frequently conceptualized as consisting of caging periods,
during which the system rattles around in a energy minimum of its energy landscape,
interrupted by transitions between minima. In this scenario
the features of the potential energy landscape (PEL) play a fundamental
role in determining the overall dynamical properties.
One way to investigate how a system explores its PEL is via the study
of the instantaneous normal modes (INM)~\cite{INM}, which are the 
instantaneous eigenvalues and eigenvectors of the Hessian matrix.
The eigenvalues give information about the curvatures
of the PEL hypersurface around the point the system is visiting, while the eigenvectors
give the corresponding directions for the joint motion of particles through phase space.
As concerns the dynamical features of the system, the negative eigenvalues
are of particular interest as they identify the unstable modes of the system~\cite{INM_D,Sciortino1997}.
Specifically, some of these negative eigenvalues are expected to be
directly related to diffusion. It is clear, however, that not all of the
unstable modes can contribute to diffusion as unstable modes are also observed at  temperatures which are sufficiently low for the diffusion constant to be negligible. 
Strategies must therefore
be devised to determine which are the unstable modes that are relevant to particle diffusion~\cite{Gezelter,LaNave2000,Clapa2012}. 

Here we explore a proposal by La Nave and coworkers~\cite{LaNave2000},
who suggested that the unstable modes allowing for diffusion are those
connecting different energy minima, known as double--well modes. 
Indeed, a one--to--one relation between the diffusion constant and the fraction $f_{\rm DW}$ of
double--well modes has been observed in a range of model systems~\cite{Li1999,LaNave2000,Barbosa2011},
We note that a different approach,
where the modes contributing to diffusion are identified with the
unstable and extended ones, and detected via a finite--size scaling analysis~\cite{Bembenek}, has recently been shown to yield analogous results~\cite{Clapa2012}.

In order to investigate the relation between the features of the potential energy landscape
and the diffusion constant of liquids, we focus on a simple model characterized by water-like density anomalies, i.e.\ regions of the phase diagram where the dynamics speeds up upon isothermal compression.
We first consider two state
points with the same diffusion constant $D$ and same temperature, but different densities,
and show that key dynamical features coincide at all length and time scales.
Indeed, not only are the diffusion constants equal, but also the mean square displacements
and the self-intermediate scattering (i.e.\ incoherent correlation) functions at different wavevectors coincide
at all times.  Differences are observed only in four--point quantities such as the non--linear dynamical susceptibility.

We then consider how the PEL changes with  density,
and how these changes are related to those in the dynamics. We show that the typical distance between two energy minima connected
by a double--well mode, as well as the associated typical energy barrier, increase monotonically with density. Both quantities 
must thus be different for two state points at different density, even if they have the same diffusion constant. 
On the other hand, the fraction of unstable normal modes
connecting different energy minima $f_{\rm DW}$ varies non--monotonically with the density. This ensures it is correlated with the diffusion constant, 
but our results do not support
the presence of a one--to--one relation between $D$ and $f_{\rm DW}$ such as has previously been seen in other systems.

\section{Model system}
As a simple model system for investigating the relation between the dynamics
and the features of the energy landscape we consider a polydisperse mixture of $N$ harmonic disks of identical mass $m$, in two dimensions. 
Diameters are uniformly distributed in the range $[d_{\rm min},d_{\rm max}]$, with
the difference $d_{\rm max}-d_{\rm  min}$ between the largest and
smallest diameter being 82\% of the mean diameter $(d_{\rm max}+d_{\rm min})/2$ so that the distribution is fairly broad; this is necessary to
prevent crystallization.
Two particles $i$ and $j$ 
interact via the repulsive harmonic potential 
\begin{equation}
 v(r) = \frac{1}{2} \epsilon \left(\frac{d-r}{d_{\rm max}}\right)^2 
 \label{eq:potential}
\end{equation}
The interaction is of finite range: particles interact only if their distance $r$ is smaller than their average diameter $d=(d_i+d_j)/2$.
Lengths, masses and energies are expressed below in units of $d_{\rm max}$, $m$ and of $\epsilon$, respectively,
and the density is specified in terms of the volume (or more precisely, area) fraction 
$\phi = N \<A\>/L^2$. Here
$L$ is the system size, $\<A\>$ the average particle area, and $N$ the
number of particles.

We have performed molecular dynamics simulations~\cite{lammps} of this
system with $N=10^3$ particles, integrating the equations of motion using
the Verlet algorithm, with a timestep of $\delta t = 10^{-4}$.
Dynamical quantities are averaged over $10^3$ independent runs. 
The system is first brought to thermal equilibrium via simulations in the NVT ensemble,
while production runs are carried out in the NVE ensemble.
Thermal equilibration is ascertained by checking for the absence of aging. 
The INMs are determined by equilibrating a larger sample, consisting of
$N = 4 \times 10^3$ particles, and averaging over $50$ independent realizations
to obtain accurate statistics.

\section{Density anomaly}
\begin{figure}[t!]
\begin{center}
\includegraphics*[scale=0.3]{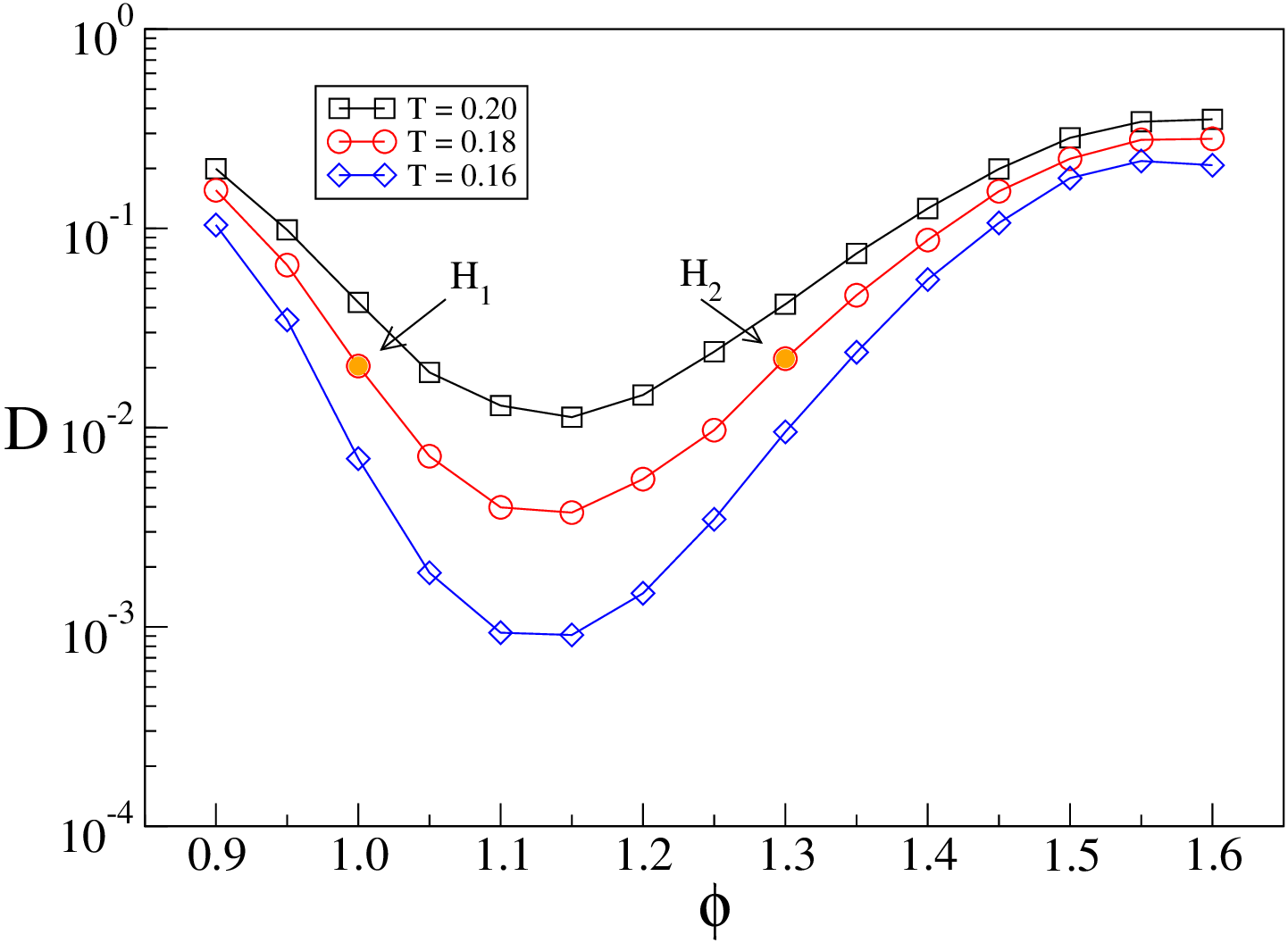}
\end{center}
\caption{\label{fig:diffusivity}
Non--monotonic density dependence of the diffusion constant. The arrows indicate the states 
$H_1 = (\phi = 1.0, T = 0.18)$ and $H_2 = (\phi = 1.3, T = 0.18)$,
which have similar diffusion constants.
}
\end{figure}
In the majority of molecular and colloidal fluids the dynamics
slows down upon isothermal compression. There are, however,
fluids that display the opposite behavior
in some region of their temperature--density phase diagram.
Such density anomalies are found in a number of fluids; the most
important of these is certainly water, with other examples including Si, Ge, Sn, and ionic melts with suitable radius ratios, 
such as SiO$_2$, BeF$_2$ and GeO$_2$~\cite{negative1,negative2}.
While all of these systems have an anisotropic interaction potential
that favors short-range tetrahedral order around the particle centers,
density anomalies have been also observed in models with
spherically symmetric interaction potentials~\cite{Buldyrev2009}. 
These systems have been studied extensively to identify the physical origin
of density anomalies and to connect them to specific features of the interaction potential.
The harmonic potential considered in this work is an example of a radially symmetric
potential giving rise to a density anomaly, as
illustrated by the non--monotonic dependence of the self--diffusion constant on volume fraction
shown in Fig.~\ref{fig:diffusivity}. We have previously investigated the physical
mechanisms responsible for this behavior, and concluded that it must be attributed
to the emergence of contacts between a particle and those of its second coordination shell~\cite{CiamarraSollich_SM,CiamarraSollich_JCP}.
In the following we first compare the dynamical features 
of two state points, $H_1 = (\phi = 1.0, T = 0.18)$ and $H_2 = (\phi = 1.3, T = 0.18)$,
that have approximately the same diffusion constant, $D_{H_2}/D_{H_1} \simeq 1.13$ (Fig.~\ref{fig:diffusivity}),
and then study the density dependence of some important features of the energy landscape.

\section{Dynamics}
\begin{figure}[t!]
\begin{center}
\includegraphics*[scale=0.3]{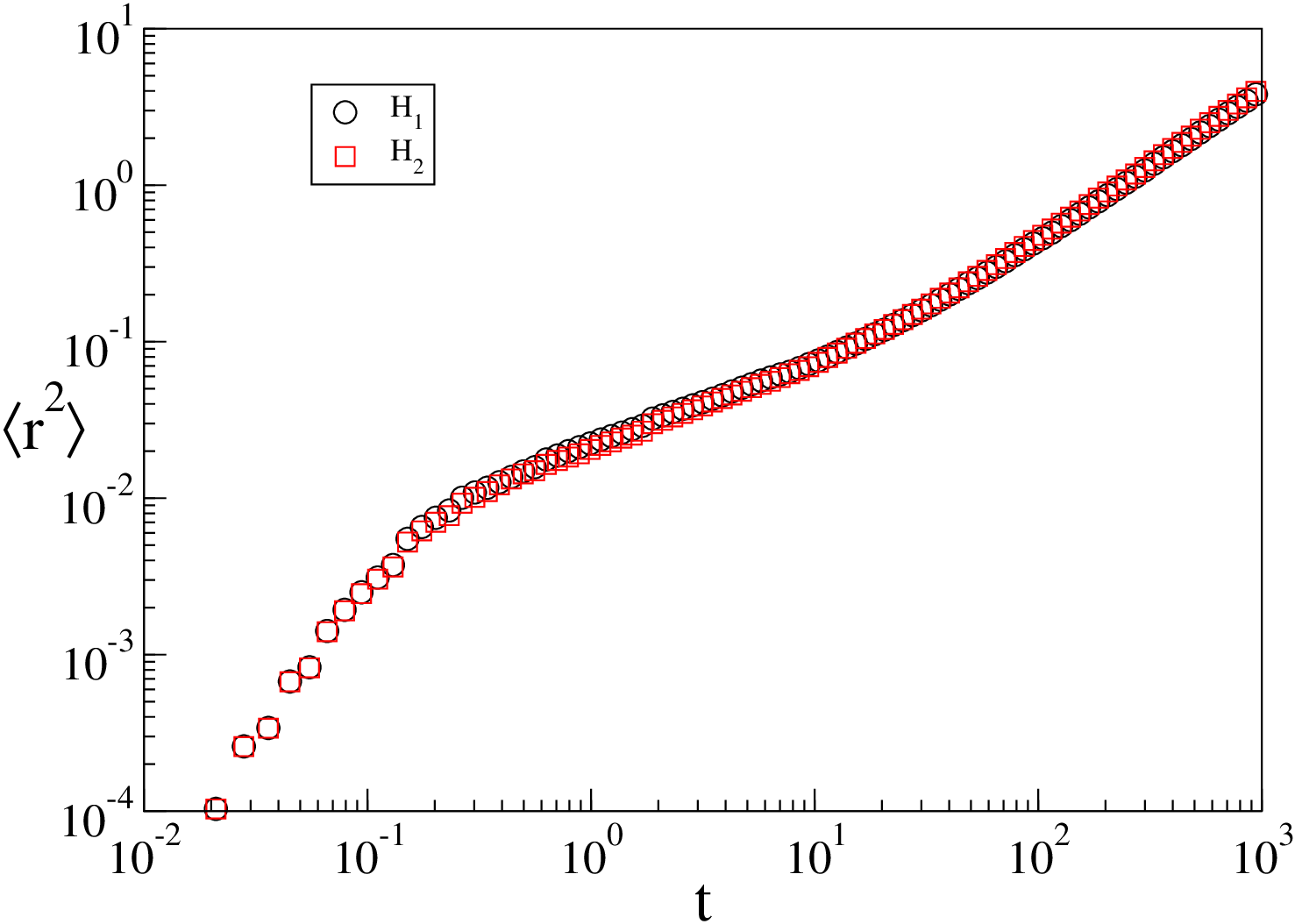}
\includegraphics*[scale=0.3]{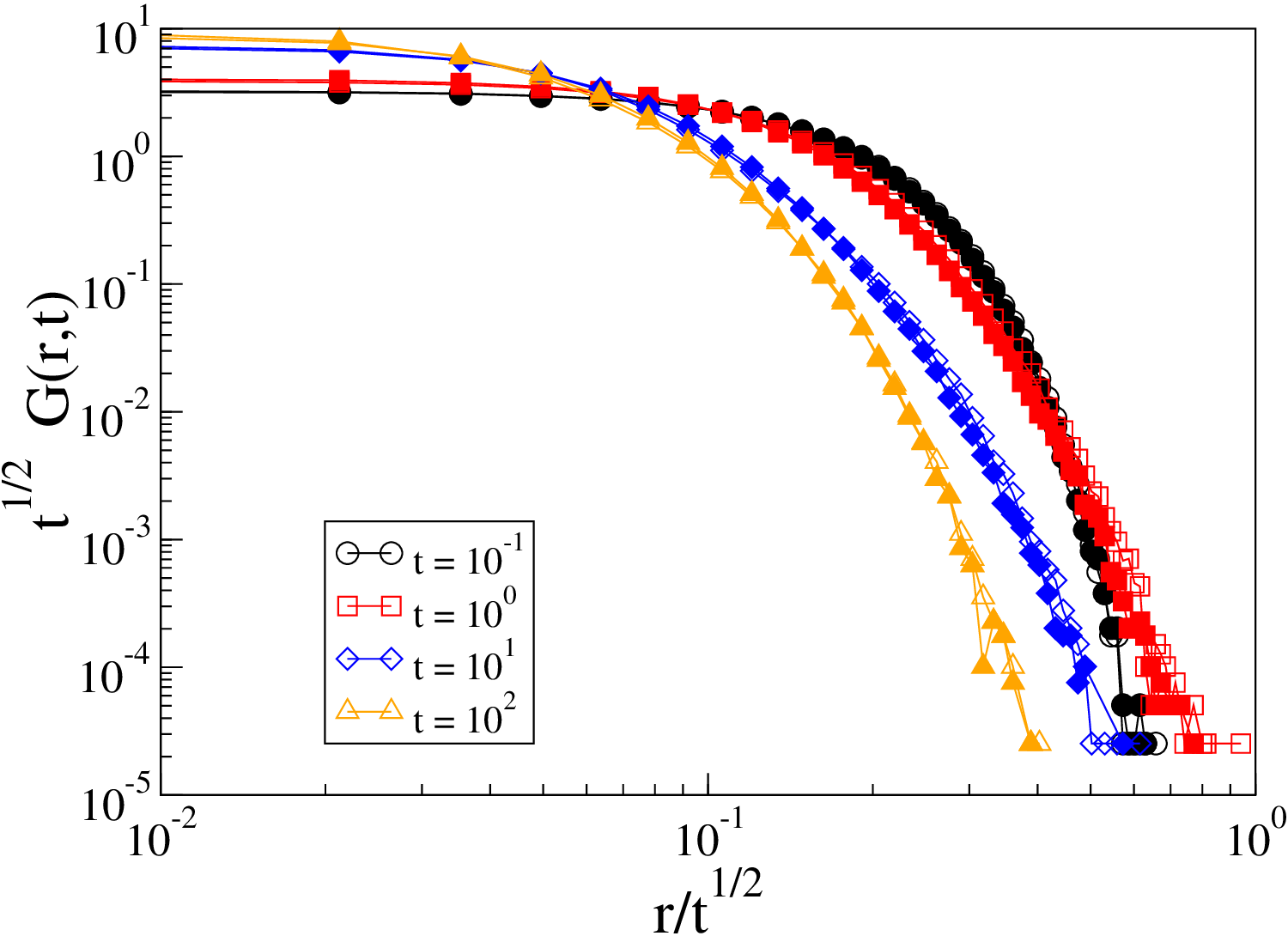}
\end{center}
\caption{\label{fig:msd}
State points $H_1$ and $H_2$ have the same mean
square displacement as a function of time (left panel), as well as
the same van Hove correlation function for every time interval $t$ (right panel),
within numerical uncertainty.
}
\end{figure}

\begin{figure}[t!]
\begin{center}
\includegraphics*[scale=0.3]{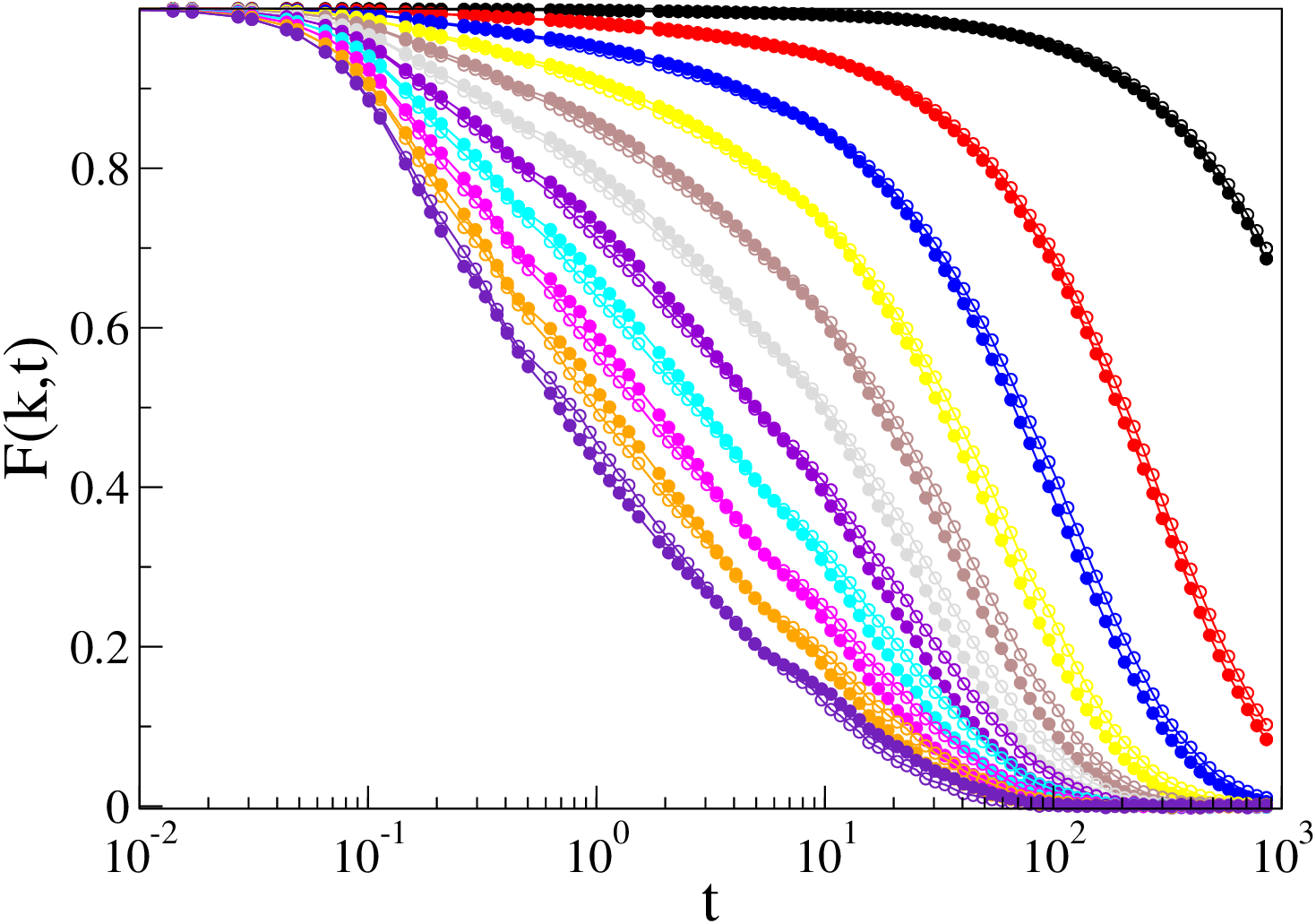}
\includegraphics*[scale=0.3]{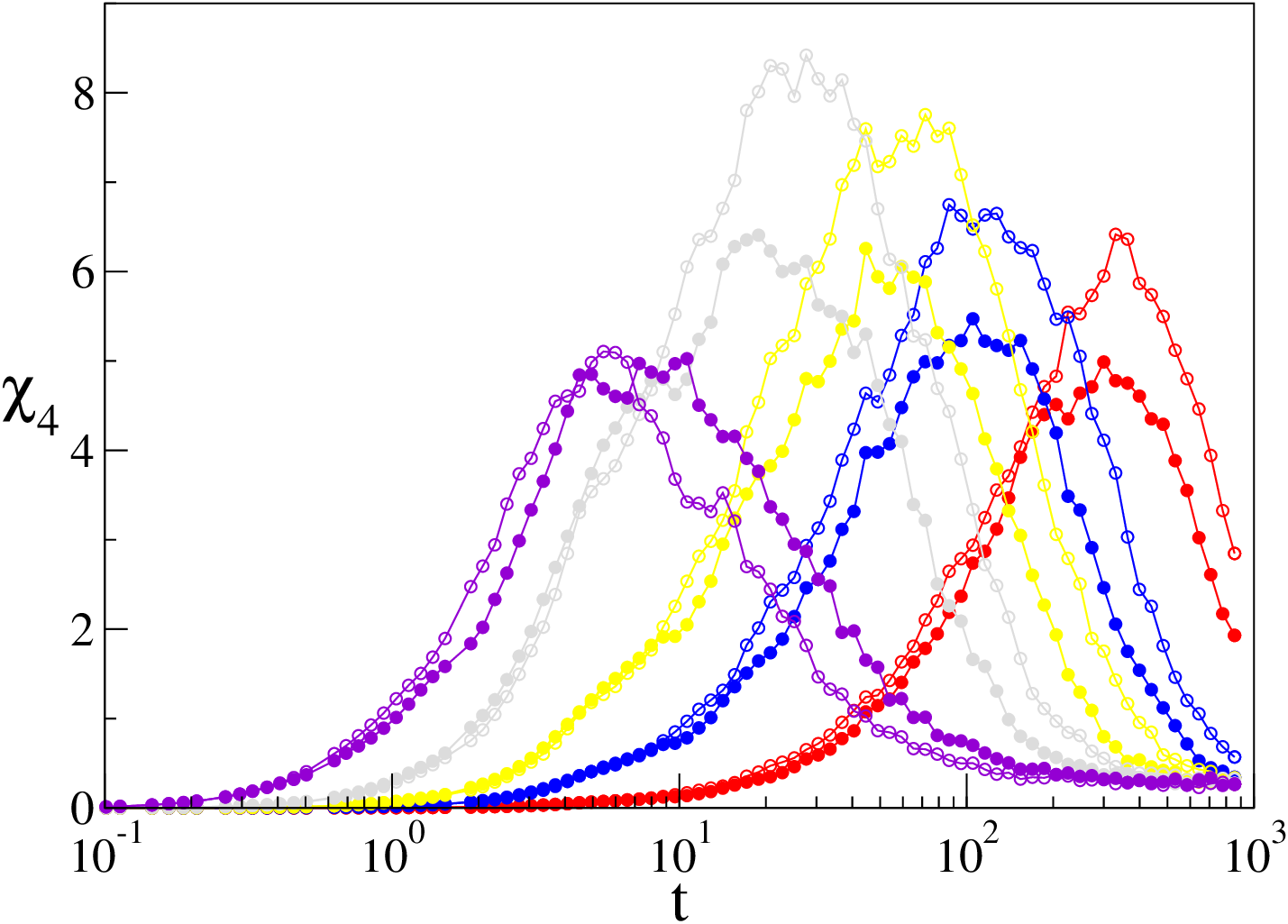}
\end{center}
\caption{\label{fig:fk}
The left panel shows the self--intermediate scattering function, $F(k,t)$, for $k = k_{\rm min} + n(k_{\rm max} - k_{\rm min})/10$,
and $n = 0,1,..,10$, from left to right. Here $k_{\rm min} = 0.2 \pi/d_{\rm max}$ and $k_{\rm max} = 4\pi/d_{\rm max}$.
Data for state point
$H_1$ (full symbols) and $H_2$ (open symbols) coincide at all times and wavevectors, to a very good approximation.
The right panel shows the corresponding dynamical suspeptibilities. For clarity, we only show data
for $n = 0,5,7,8,9$, from left ro rigth.
}
\end{figure}

As illustrated in Fig.~\ref{fig:diffusivity}, the presence of a
density anomaly means that at a given temperature there are pairs of
points that have different density but the same diffusion constant, such as state points $H_1$ and $H_2$.
Due to the difference in density, one might expect the dynamics at
these state points to be different in general, apart from sharing an
overall timescale set by the diffusion constant. Surprisingly, we find that
the opposite is true: state points with the same
diffusion constant share many dynamical features.
Consider first the mean square displacement shown in
Fig.~\ref{fig:msd}. It is equal between the two state points not only 
at long time, where this is expected because the two states have equal
diffusion constant, but in fact at all times to within our numerical accuracy.
This is consistent with the existence of a close link between
the Debye--Waller factor, which is the value of the mean square displacement
in the plateau region, and the
diffusion constant. Such a link has been put forward on more general
grounds~\cite{Dyre06}, and we have previously verified that it applies
across a large part of the phase diagram of the model system considered here~\cite{CiamarraSollich_JNCS}. 
The mean square displacement is given by the variance of the
self part of the van Hove function, $G(r,t)$, defined here as the probability that a
particle has moved a distance $r$ along one coordinate (say $x$-)axis at time $t$. 
To investigate the relation between the dynamics at state points $H_1$ and $H_2$ more closely, we therefore next look at the van Hove function itself.
Fig.~\ref{fig:msd} shows that, surprisingly, the van Hove functions --- evaluated for the same time interval $t$ -- at
$H_1$ and $H_2$ coincide not just in their variance, but in their entire functional dependence on $r$. 
The observed equality of the van Hove functions suggests that in fact the dynamics of the two systems might be equal at all relevant length scales. We have checked this by measuring 
the self--intermediate scattering (or incoherent correlation) function,
$
 F(k,t) = \frac{1}{N} \left\langle \sum_{j=1}^N \exp\left[ -i \bm{k}\cdot (\bm{r}_j(t) -\bm{r}_j(0)) \right] \right\rangle
$
for $11$ different wavevectors $\bm{k}$ with lengths $k=|\bm{k}|$ evenly spaced between $k_{\rm max} = 4\pi/d_{\rm max}$ and $k_{\rm min} = 0.2 \pi/d_{\rm max}$. 
As shown in Fig.~\ref{fig:fk}, $H_1$ and $H_2$ have very similar self--intermediate scattering functions,
at all times and length scales. 
Of course $F(k,t)$ is related to $G(r,t)$, essentially by Fourier transform, so this equality primarily provides a consistency check that equality of the dynamics 
is found both in real and Fourier space. We note that a similar collapse cannot be observed in the distinct part of the
van Hove function, $G_d(r,t)$, as at $t = 0$ this coincides with the radial distribution function, which is
distinct for states having different densities. Similarly, no collapse can be observed in the total--intermediate
scattering function.

Digging deeper, differences in the dynamics at state points $H_1$ and $H_2$ finally turn up in the {\em fluctuations}  of the self--intermediate scattering function. If we denote this fluctuating version of $F(k,t)$ by 
$
\mathcal{F}(k,t) = \frac{1}{N} \sum_{j=1}^N \exp\left[ -i \bm{k}\cdot (\bm{r}_j(t) -\bm{r}_j(0)) \right]
$
then its variance defines the so--called dynamical susceptibility $\chi(k,t) = N [ \<\mathcal{F}(k,t)\>^2 - \<\mathcal{F}(k,t)^2\>]$. 
In this quantity we do find clear differences between $H_1$ and $H_2$, as shown in Fig.~\ref{fig:fk}. In particular, while 
the susceptibilities attain their maximum roughly at the same time
$t^*$ for any given $k$, 
the actual maximal value $\chi(k,t^*)$ is smaller for state $H_2$, i.e.\ in the anomalous region.

\section{Instantaneous normal--modes}
Above we have found strong similarities between the dynamical properties of the equal-diffusion constant state points $H_1$ and $H_2$.
If there is a significant correlation between the dynamics and the properties of the energy landscape, then this would suggest
that the energy landscapes of states $H_1$ and $H_2$ should also have closely related features. More specifically this should apply to the statistics of those landscape properties that are directly related to the dynamics. 
Accordingly, by comparing different features of the energy landscape of the two states,
we can infer to what extent they correlate with the dynamics.

\begin{figure}[t!]
\begin{center}
\includegraphics*[scale=0.3]{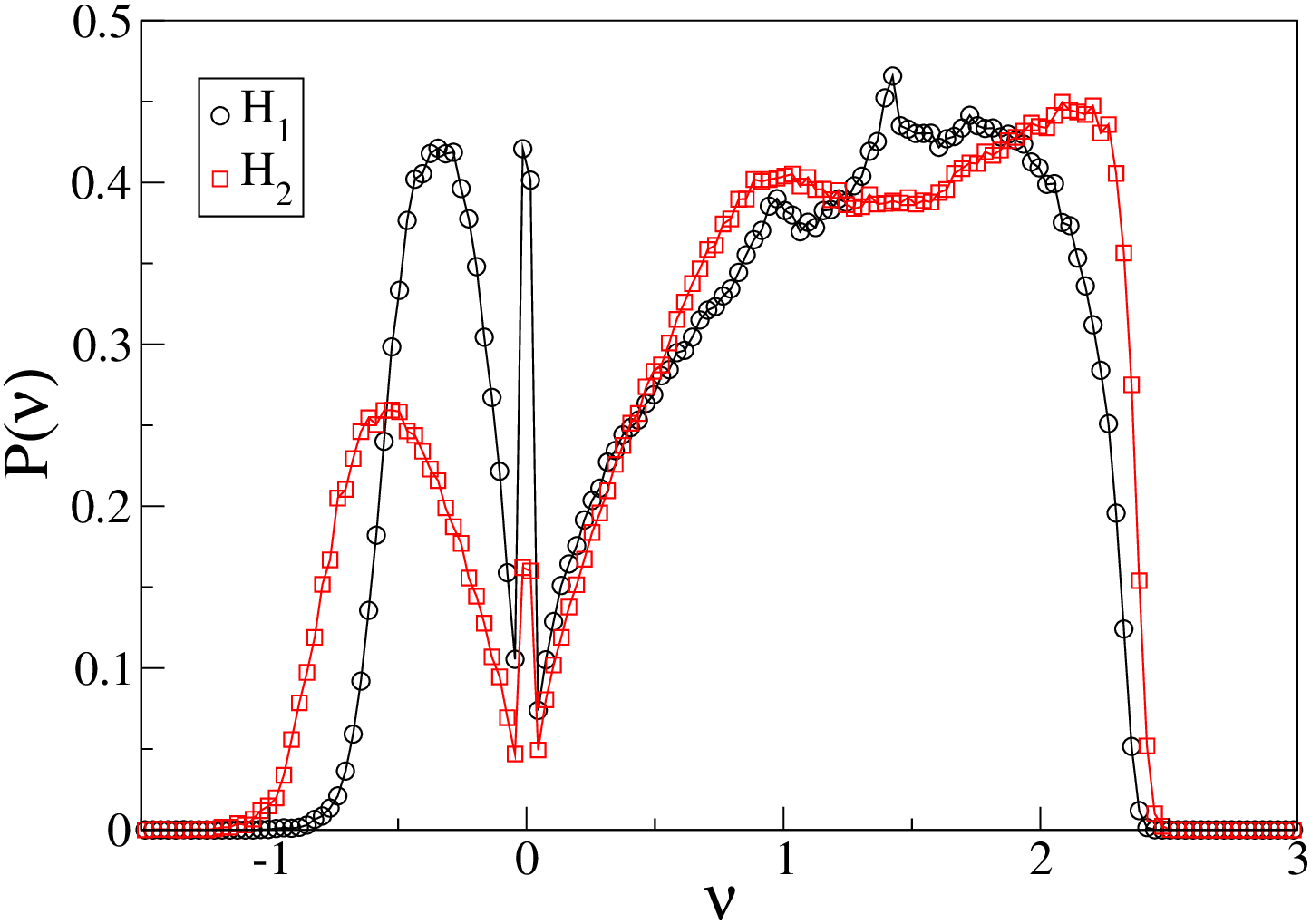}
\end{center}
\caption{\label{fig:dnu}
Probability distribution of the eigenfrequencies of state points $H_1$ and $H_2$, for a system of $N=4,000$ particles. 
Imaginary eigenfrequencies $\nu$ are plotted as $-|\nu|$ to separate them from the real eigenfrequencies.
}
\end{figure}

To characterize the energy landscape statistics we employ the instantaneous normal mode approach, and thus focus on the properties of the eigenvalues $\lambda_i$
and the eigenvectors $\bm{\delta u}_i$ of the Hessian matrix of the system. 
We consider first the distribution $P(\nu)$ of the eigenfrequencies $\nu_i = \sqrt{\lambda_i}$. These are shown in Fig.~\ref{fig:dnu}, where following the convention in the literature we plot imaginary eigenfrequences arising from negative eigenvalues $\lambda_i$ as $-|\nu_i|=-\sqrt{|\lambda_i|}$. They can then be plotted on the same axis as the real eigenfrequencies, but remain distinct from them in such a visualization. 
Fig.~\ref{fig:dnu} demonstrates clearly that the resulting two eigenvalue distributions differ between the state points $H_1$ and $H_2$. This is presumably driven by the difference in density between the two states.

\begin{figure}[t!]
\begin{center}
\includegraphics*[scale=0.3]{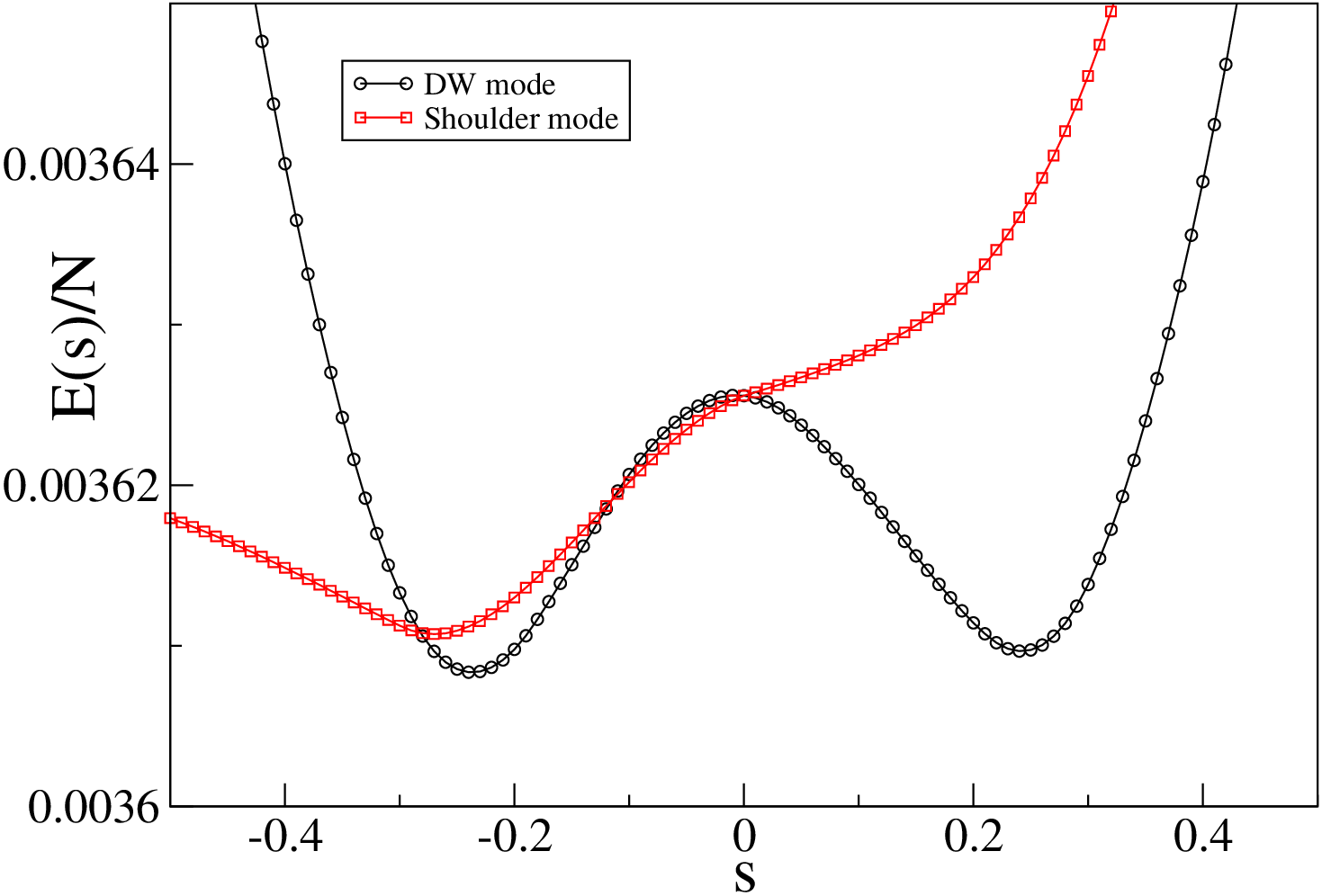}
\end{center}
\caption{\label{fig:Ei}
For each unstable eigenvalue $i$ we have displaced the particle positions by an amounty $s\,\bm{\delta u}_i$ along the 
eigenvector $\bm{\delta u}_i$; the eigenvector is normalized so that $s^2$ gives the total squared displacement of all particles. The dependence of the energy of the system $E$ on $s$ allows one to discriminate between double--well modes and shoulder modes,
as illustrated in the figure for two typical cases. The figure refers to state $H_1$ and $N = 4,000$.
}
\end{figure}

The diffusive dynamics of the system, however, is not influenced by all modes.
La Nave and coworkers~\cite{LaNave2000}
suggested that the unstable modes allowing for diffusion are the
double--well modes, i.e.\ those connecting different energy minima, and 
observed a one--to--one relation between the diffusion constant $D$ and the fraction of modes, $f_{DW}$, that have this double--well character.
In order to determine if an unstable mode $i$ is a double--well mode, one asks 
how the interaction
energy of the system $E_i(s)$ changes as particles are displaced by
$s\, \bm{\delta u}_i$ along the eigenvector. Note that, since we consider unit eigenvectors,
$s$ measures the distance the system is displaced in phase space: $s^2$ is the sum of all squared particle displacements.
If $E_i(s)$ is a double--well function, i.e.\ has two local minima, then we classify the mode
$i$ as a double--well mode. As an illustration, we show in Fig.~\ref{fig:Ei}
numerical results for $E_i(s)$, for the two cases of a double--well and of a ``shoulder'' mode.
We note that all modes with approximately zero eigenvalue, which are responsible for the observed peak
in $P(\nu)$ around $\nu = 0$, turn out to be shoulder modes.

\begin{figure}[t!]
\begin{center}
\includegraphics*[scale=0.3]{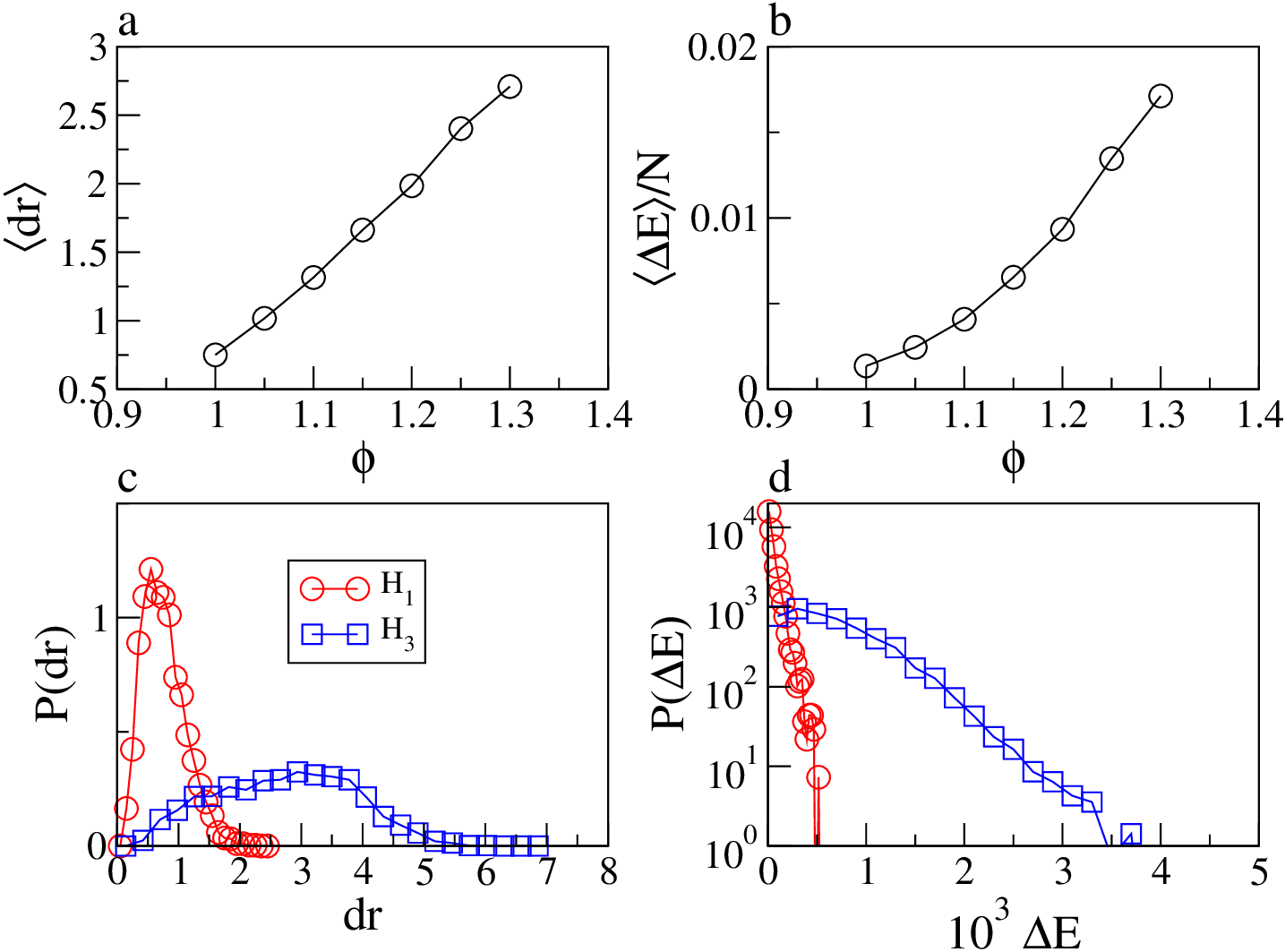}
\end{center}
\caption{\label{fig:transition_density}
Bottom row: Probability distribution of the distance $dr$ between two energy minima connected
by a double--well mode (bottom left), and of the energy barrier ${\Delta E}$ separating the two minima (bottom right), for the two states $H_1$ and $H_2$. 
Top row: averages of these quantities as a function of volume fraction, across the range between $H_1$ ($\phi=1$) and $H_2$ ($\phi=1.3$). 
}
\end{figure}

The features of the double--well modes change with density. As displayed in
Fig.~\ref{fig:transition_density}, the average distance $\<dr\>$ separating two energy minima in phase space
and the corresponding average energy barrier $\<\Delta E\>$, increase monotonically with density. In particular, these quantities are therefore not identical for the two states $H_1$ and $H_2$, suggesting that they are at most weakly correlated with the dynamics. 
The figure (bottom row) also show that $dr$ is approximately Gaussian
distributed, while $\Delta E$ has a roughly exponential distribution. 
We already know the averages of the two distributions are different for $H_1$ and $H_2$, and accordingly also the distributions themselves are quite distinct.

\begin{figure}[t!]
\begin{center}
\includegraphics*[scale=0.3]{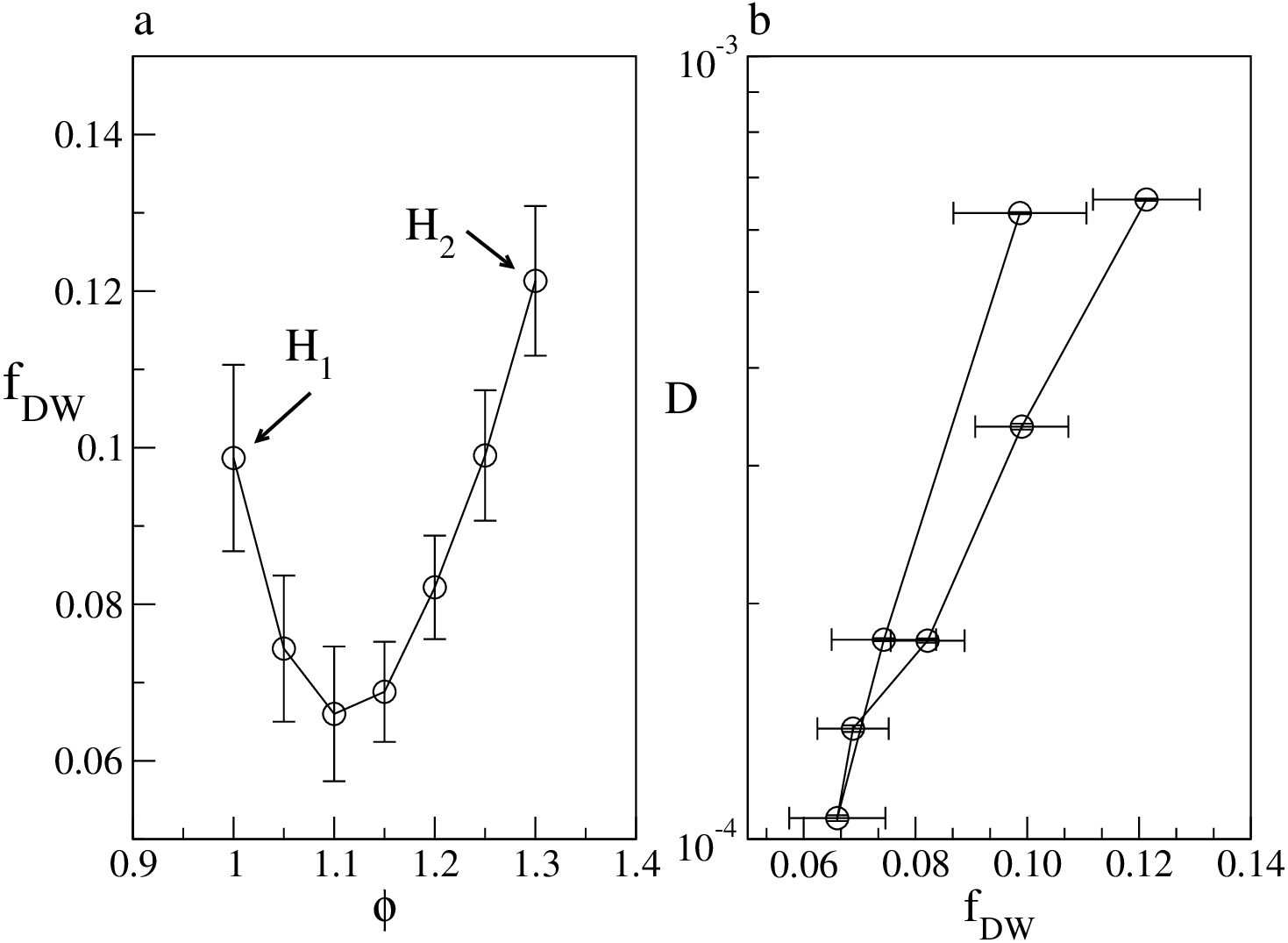}
\end{center}
\caption{\label{fig:fdw}
(a) The fraction $f_{\rm DW}$ of unstable modes connecting different energy minima, i.e.\ double-well modes, varies non--monotonically with the volume fraction. (b) It is therefore strongly correlated with the diffusion constant, but does not determine it uniquely.
}
\end{figure}

So far we have found for the instantaneous normal mode eigenfrequency distribution, 
and geometrical properties of double--well modes such as distance and barrier between the two minima, 
that states $H_1$ and $H_2$ show clear differences in their energy landscape statistics. 
We now turn, finally, to the fraction $f_{DW}$ of double--well modes. As shown in Fig.~\ref{fig:fdw}a, 
this varies non--monotonically with the volume fraction, much like the diffusion constant. 
As expected, a parametric plot of $D$ versus $f_{DW}$ in Fig.~\ref{fig:fdw}b therefore exhibits a strong correlation between these two quantities. 
The fraction of double--well modes is therefore clearly one feature of the energy landscape that is closely linked to the dynamics, 
and indeed (Fig~\ref{fig:fdw}a) its values in states $H_1$ and $H_2$ are within $\approx 23\%$ of each other. 
However, Fig~\ref{fig:fdw}b also demonstrates that for our system there is no one--to--one correspondence between $D$ and $f_{DW}$, 
in contrast to what has been observed in other systems~\cite{LaNave2000}.

\section{Conclusions}
We have investigated a polydisperse mixture of disks interacting via
a repulsive finite-range harmonic potential, in the volume fraction and temperature range where the system
displays a density anomaly. We have focussed primarily on the possible existence of correlations
between the dynamics and features of the potential energy landscape.
We have observed striking similarities between the dynamical features
of state points characterized by the same diffusion constant: as far as can be ascertained from incoherent (i.e.\ single particle) two-point correlations, their dynamics
appear to coincide at all length and time scales. We note that these dynamical 
similarities are not confined to the two specific state points $H_1 = (\phi=1, T = 0.18$) and $H_2 = (\phi=1.3, T = 0.18)$ that we have studied; we have obtained analogous results when comparing e.g.\ the dynamics at the state points $(\phi=1.03, T = 0.16$) and $(\phi=1.25, T = 0.16)$.
What is the physical origin of these dynamical similarities is not obvious to us, and will be an interesting question to follow up in future work.
In this respect, we mention also that not all liquids with density anomalies have this feature; indeed, for a repulsive Gaussian interaction we generally find quite distinct dynamics for states of equal diffusion constant~\cite{Gaussian}.

Despite the dynamical 
similarities we have found in our specific model system, the states that are paired up by having equal diffusion constant have PELs with quite different statistics. These include 
the average distance between minima connected by unstable normal modes, and the average barrier. One quantity where we do find a strong correlation with the dynamics is in the fraction $f_{\rm DW}$ of double--well modes among all instantaneous normal modes; such double--well modes are defined as unstable modes connecting different energy minima. 
However, while there is a clear correlation between $f_{\rm DW}$ and $D$, the two quantities are not uniquely determined by each other. This suggests that in our system there must be other features of the landscape, beyond $f_{\rm DW}$, that act to determine the diffusion constant $D$.

\end{document}